\documentclass[12pt]{article} 

\usepackage{times} 
\usepackage{wrapfig,lipsum,booktabs,amssymb,amsmath}
\usepackage{setspace}
\usepackage{bm} 
\usepackage{color} 
\usepackage{hyperref} 
\usepackage{xr-hyper}
\usepackage[font=small, belowskip=-14pt,aboveskip=0pt]{caption}
\usepackage{graphicx} 
\usepackage{vmargin} 
\usepackage{titlesec}
\usepackage{multirow}
\usepackage[normalem]{ulem}
\usepackage{braket}
\usepackage{cancel}
\usepackage{makecell}
\titlespacing{\section}{5pt}{\parskip}{-\parskip}
\titlespacing{\subsection}{3pt}{\parskip}{-\parskip}
\titlespacing{\subsubsection}{2pt}{\parskip}{-\parskip}
 
\usepackage[style=nature,citestyle=numeric-comp, sorting=none, backend=biber, maxbibnames=9]{biblatex}
\hypersetup{colorlinks=true, citecolor=black,linkcolor=blue,urlcolor=blue}
\DeclareCaptionFont{blue}{\color{blue}}
\usepackage{etoolbox}
\patchcmd{\thebibliography}{\section*{\refname}}{}{}{}
\usepackage{authblk}

\setpapersize{USletter}
\setmarginsrb{1in}{1in}{1in}{1in}{0pt}{0mm}{0pt}{0mm}

\begin{document}

\title{AI-Assisted Identification of Magnetic Orders and Skyrmions}

\author[1]{Haowen Yang}
\author[1]{Sophia Huerta}
\author[1]{Yingying Wu\footnote{Corresponding author: yingyingwu@ufl.edu}}

\affil[1]{Department of Electrical and Computer Engineering, University of Florida, Gainesville, FL 32611, USA}

\date{\vspace{-5ex}}
\maketitle
\begin{abstract}
Exotic magnetic orders in two-dimensional (2D) materials are attracting huge interest for energy-efficient spintronic applications, yet realizing robust high-temperature van der Waals antiferromagnets and topological magnetic states remains challenging. In this work, we develop a machine-learning framework for identifying magnetic orders and predicting magnetization using structural, compositional, and electronic information derived from the Materials Project. Fixed-length descriptors are constructed for two complementary tasks: ferromagnetic (FM) versus antiferromagnetic (AFM) classification and quantitative magnetization prediction. Magnetic order is classified using a LightGBM model trained on structure-derived descriptors without explicitly including magnetic descriptors. Five-fold cross-validation grouped by chemical system is used to reduce chemical leakage, together with hyperparameter and classification-threshold optimization. On an isolated test set, the classifier achieves a balanced accuracy of $93.6\%$ in the testing set. Magnetization is predicted using a residual multilayer perceptron with compositional, structural, electronic, and task-specific initial-state descriptors. A logarithmic target transformation and robust weighted loss are used to account for the broad magnetization distribution, while three independently trained models are combined into a final ensemble. The model achieves a mean absolute error of $0.686~\mu_{\mathrm{B}}$/formula unit. For selected magnetic candidates, simulated magnetic imaging and phase-reconstruction analysis are further used to investigate magnetization textures and skyrmion-like features through normalized magnetization profiles and topological charge. This framework provides an efficient approach for screening 2D magnetic materials and prioritizing candidates for antiferromagnetic and topological spintronic applications.
\end{abstract}

\section*{Introduction}
Magnetic skyrmions are nanoscale, topologically protected spin textures with a swirling real-space magnetization configuration. Their topology is characterized by a winding number that describes how the local spin texture wraps around the order-parameter space \cite{zhang2017direct}. Due to their topological stability, nanoscale dimensions, and particle-like dynamics, skyrmions have attracted significant interest for future spintronic, neuromorphic, and quantum technologies \cite{zhang2017direct, tokura2020magnetic,wu2020neel,wu2022van,zhang20242d,yang2026skyrmion,zhong2025integrating,plummer20252d,wu2025spin,sosa2025simulating,wu2024room,yang2026skyrmion,zhong2025integrating}. Recent studies have demonstrated the new functions of skyrmion-based systems, all the way from electrically detectable magnetic states to unconventional information-processing and high-fidelity quantum circuits \cite{wang2020topological,yang2026skyrmion}. These developments emphasize nonconventional magnetic or quantum devices, and require materials with preferred magnetic properties, such as room-temperature magnetization in Fe$_3$GaTe$_2$ layers \cite{zhang2022above}, strong exchange interactions at ferromagnet/antiferromagnet heterostructure \cite{wu2022manipulating}, and energy-efficient multiferroic control using CuInP$_2$S$_6$/Fe$_3$GaTe$_2$ interface \cite{wu2024room}. Efficient prediction and screening of these fundamental magnetic properties therefore represent an important step toward discovering materials suitable for future technologies. 

Additionally, 2D van der Waals magnets provide a powerful platform for engineering such spin textures because their weak interlayer bonding enables exfoliation, stacking, twisting, gating, and interface design \cite{zhang20242d,rahman2021recent,zhong2025integrating,wu2019induced,wang2020topological}. Since the discovery of intrinsic magnetism in atomically thin materials, 2D magnets have expanded their functionalities by controlling magnetic anisotropy, exchange coupling, Dzyaloshinskii–Moriya interaction, and interfacial spin-orbit effects \cite{zhang20242d,rahman2021recent}. Recent studies have shown that van der Waals magnetic heterostructures can stabilize nanoscale skyrmions and enable their detection through magnetic imaging and transport measurements such as the topological Hall effect \cite{wu2020neel,wu2022van, zhang20242d}. These results suggest that interface engineering in layered magnetic materials is crucial for designing skyrmion-hosting systems for future spintronics applications.
The traditional computational discovery of magnetic materials often relies on density functional theory (DFT) and related first-principles calculations. DFT has been central to predicting band structures, formation energies, magnetic anisotropy, exchange interactions, and other properties of candidate materials. Large databases such as Materials Project, C2DB, and MC2D have made DFT-derived material properties accessible for high-throughput screening \cite{sodequist2024magnetic, haastrup2018computational, campi2023expansion, faber2017prediction, del2023deep, jain2013commentary}. 
With the increasing availability of large and diverse materials datasets, machine learning (ML) methodologies have progressively evolved toward deep-learning architectures capable of learning physically meaningful representations directly from atomic structures. In particular, crystal graph convolutional neural networks and related graph-based models represent crystalline materials through interconnected atomic environments, thus reducing dependence on predefined descriptors and enabling direct learning of complex structure–property relationships \cite{xie2018crystal,chen2019graph}. Beyond computational property prediction, ML has also become increasingly important for experimental materials characterization. For example, ML is used for facilitating materials thickness characterization \cite{leger2024machine}. More broadly, ML has been integrated with DFT, high-throughput computational screening, and experimental characterization to accelerate the exploration of large material spaces and reduce the computational and experimental resources needed for material discovery \cite{ramprasad2017machine,sorkun2020artificial}. Recent studies have also demonstrated that ML provides a complementary strategy by learning patterns from existing DFT or experimental data and rapidly predicting material properties for new candidates \cite{faber2017prediction,del2023deep,jain2013commentary,rhone2020data,kabiraj2020high}.

In this work, we use the ML-based approach to identify and characterize potential 2D magnets using data obtained from the Materials Project without imposing a specific crystal prototype or substitution pattern.
Structural information and electronic band structures are converted into physically motivated numerical descriptors that capture characteristics related to crystal symmetry, chemical composition, electronic states, spin polarization, and magnetic behavior.
Two complementary prediction models are developed. The first model identifies the magnetic ordering of a material by classifying candidates as FM or AFM, while the second model predicts the continuous magnetization value of each material. Because these tasks require different learning objectives, separate machine-learning strategies are used for classification and regression. These two models provide both qualitative information about magnetic ordering and quantitative information about magnetic strength, allowing candidate 2D AFM materials to be screened more efficiently than by performing additional first-principles calculations for every material. This approach follows the broader direction of high-throughput and machine-learning-assisted materials discovery, in which existing computational databases are used to accelerate the exploration of large chemical and structural spaces. 
More importantly, this work also explores a ML-enabled framework for topological order identification in 2D magnets. The developed workflow focuses on processing magnetic or optical image data to recover information related to spin textures and phase structure. Techniques such as Lorentz transmission electron microscopy and transport-of-intensity equation (TIE) analysis have been used to reconstruct magnetic induction maps and interpret complex spin textures.

\section*{Results}

\subsection*{Magnetic Order Classification}

\begin{figure}[ht!]
    \centering
    \includegraphics[width=\linewidth]{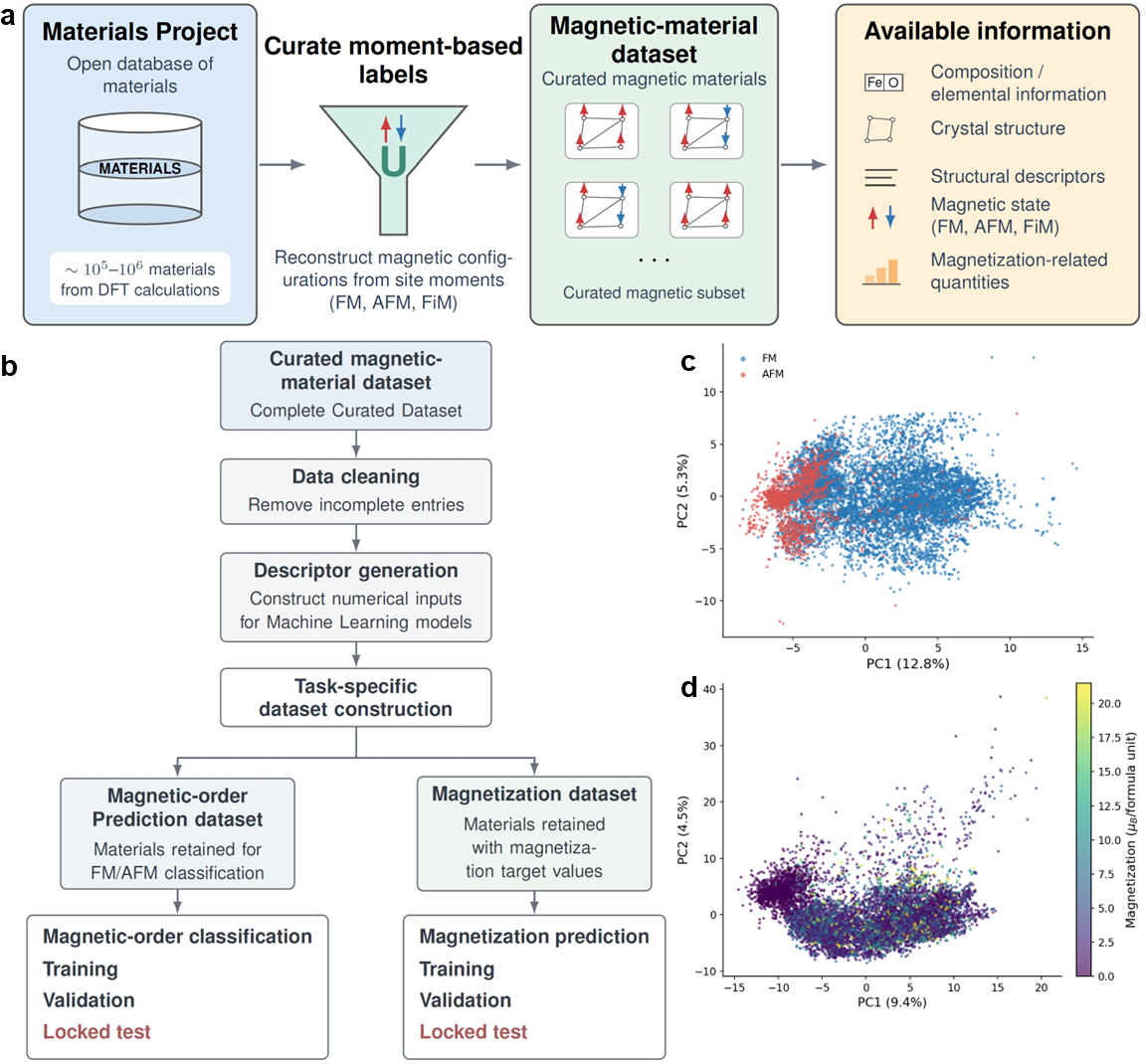}
    \caption{Dataset curation and pre-processing pipeline and feature-space visualization.
    \textbf{(a)} Overview of the magnetic-material data acquisition process. Materials are retrieved from the Materials Project, and magnetic configurations are reconstructed from available site-resolved moments. Missing task-level moments are supplemented by targeted material-structure recovery for Materials Project AFM and FiM candidates. The resulting curated magnetic-material dataset includes compositional, structural, magnetic-state, and magnetization-related information used for subsequent machine-learning tasks.
    \textbf{(b)} Data-processing and task-specific dataset construction workflow. The curated magnetic dataset is cleaned to remove incomplete entries and transformed into numerical descriptors suitable for machine-learning models. Separate datasets are then constructed for magnetic-order classification and magnetization prediction, with each task divided into training, validation, and locked test sets.
    \textbf{(c)} Principal component analysis (PCA) projection of the descriptor space for the magnetic-order classification dataset, showing the distributions of FM and AFM materials. The first two principal components account for 12.8\% and 5.3\% of the total variance, respectively.
    \textbf{(d)} PCA projection of the magnetization-prediction dataset, with each material colored according to its magnetization in $\mu_{\mathrm{B}}$ per formula unit. The first and second principal components account for 9.4\% and 4.5\% of the total variance, respectively.}
    \label{fig:1}
\end{figure}

Materials data were retrieved from the Materials Project using the \texttt{mp-api} interface through \texttt{MPRester} \cite{jain2013commentary}. The construction and physics-informed curation of the magnetic-material dataset are summarized in Fig.~\ref{fig:1}. 
The initial query returned 271,325 materials with Materials Project magnetic-order information, consisting of 150,126 nonmagnetic (NM), 74,947 ferromagnetic (FM), 16,759 ferrimagnetic (FiM), 3,512 antiferromagnetic (AFM), and 25,981 materials with an unknown magnetic ordering. Rather than using these database-assigned orderings directly as machine-learning targets, the corresponding source calculations were traced through the Materials Project \texttt{origins} records. For each material, the task associated with the ``magnetism'' origin was identified and queried through the Materials Project task endpoint. The final site-resolved magnetic moments were extracted from the \texttt{magmom} site property of the relaxed task structure, while the initial moments specified in the calculation input were retained only for diagnostic purposes. Consequently, the target labels used in this work were reconstructed from the final DFT magnetic configuration rather than from the initial spin initialization or the Materials Project categorical magnetic label.

This distinction is physically important because magnetic order is determined by the arrangement of \emph{local magnetic moments}, rather than by the total magnetic moment of the unit cell alone. Within the collinear treatment used for these calculations, the moment on each atomic site can be represented by a signed scalar $m_i$, where the sign describes the relative spin orientation. A ferromagnetic configuration is characterized by local moments aligned predominantly in the same direction, whereas an antiferromagnetic configuration contains finite oppositely aligned moments whose contributions approximately compensate. A ferrimagnetic state also contains oppositely oriented magnetic sublattices, but their unequal magnitudes result in incomplete compensation and a finite residual moment. Thus, a small total magnetic moment is not sufficient to identify an AFM state, since both a compensated antiferromagnet and a genuinely nonmagnetic material may exhibit $M_{\mathrm{tot}}\approx0$ \cite{horton2019high}.

To distinguish these cases, the magnetic-state reconstruction was performed using the site-resolved final moments with NumPy-based numerical analysis. Weak site moments were first removed by defining the set of magnetically active sites as

\begin{equation}
    \mathcal{A}
    =
    \left\{
    i:\left|m_i\right|\geq m_{\mathrm{th}}
    \right\},
\end{equation}
where the baseline active-moment threshold was
$m_{\mathrm{th}}=0.20~\mu_{\mathrm{B}}$. This cutoff suppresses weak residual spin polarization that may arise numerically in a self-consistent DFT calculation. If no sites remained in $\mathcal{A}$, the configuration was classified as NM. For the remaining active moments, the absolute and residual magnetic moments were defined as

\begin{equation}
    M_{\mathrm{abs}}
    =
    \sum_{i\in\mathcal{A}}\left|m_i\right|,
    \qquad
    M_{\mathrm{net}}
    =
    \left|
    \sum_{i\in\mathcal{A}}m_i
    \right|.
\end{equation}

A normalized alignment parameter was then introduced as

\begin{equation}
    A
    =
    \frac{M_{\mathrm{net}}}{M_{\mathrm{abs}}},
\end{equation}
together with the corresponding cancellation parameter

\begin{equation}
    C = 1-A.
\end{equation}

For moments aligned predominantly in one direction, $A\rightarrow1$ and $C\rightarrow0$, whereas strongly compensated configurations approach $A\rightarrow0$ and $C\rightarrow1$. This normalization is particularly useful because it measures the degree of magnetic compensation independently of the absolute magnitude of the local moments.

The positive and negative active magnetic sublattices were additionally evaluated separately according to

\begin{equation}
    M_{+}
    =
    \sum_{\substack{i\in\mathcal{A}\\m_i>0}}m_i,
    \qquad
    M_{-}
    =
    \left|
    \sum_{\substack{i\in\mathcal{A}\\m_i<0}}m_i
    \right|,
\end{equation}
and their relative balance was quantified as

\begin{equation}
    B
    =
    \frac{\min(M_{+},M_{-})}
         {\max(M_{+},M_{-})}.
\end{equation}

Here, $B\rightarrow1$ corresponds to two nearly equal and oppositely oriented magnetic sublattices, whereas smaller values indicate progressively stronger sublattice imbalance. The numbers of positive and negative active sites, $N_{+}$ and $N_{-}$, were also explicitly counted. The baseline magnetic-state assignment implemented in the curation procedure can therefore be written as \cite{merker2022machine}

\begin{equation}
L =
\begin{cases}
\mathrm{NM},
    & N_{\mathrm{active}}=0, \\[3pt]
\mathrm{FM},
    & N_{+}=0\ \mathrm{or}\ N_{-}=0, \\[3pt]
\mathrm{AFM},
    & A\leq0.05\ \mathrm{and}\ B\geq0.90, \\[3pt]
\mathrm{FiM},
    & \mathrm{otherwise}.
\end{cases}
\end{equation}

Thus, an AFM label required not only a small residual moment relative to the total local-moment magnitude, but also two oppositely oriented magnetic sublattices with at least 90\% balance. This additional balance requirement prevents a strongly uncompensated FiM configuration from being identified as AFM solely because its net moment is small.

Because the distinction between weak local magnetism, compensated AFM order, and FiM order can depend on numerical thresholds, the stability of each assignment was explicitly tested rather than relying on a single cutoff. The labeling procedure was repeated for

\begin{equation}
    m_{\mathrm{th}}
    \in
    \left\{
    0.10,\,
    0.15,\,
    0.20,\,
    0.25,\,
    0.30
    \right\}
    ~\mu_{\mathrm{B}},
\end{equation}
and

\begin{equation}
    A_{\max}
    \in
    \left\{
    0.02,\,
    0.05,\,
    0.10
    \right\},
\end{equation}
while maintaining $B_{\min}=0.90$. The resulting $5\times3=15$ threshold combinations were compared with the baseline assignment obtained using
$m_{\mathrm{th}}=0.20~\mu_{\mathrm{B}}$ and $A_{\max}=0.05$ as discussed in Supplementary Information Section \ref{sec:supp_thressentivity}. The labeling criterion identifies robust compensated collinear magnetic configurations rather than uniquely determining the microscopic antiferromagnetic ordering pattern. For a baseline label $L_0$, the label robustness was defined as

\begin{equation}
    R
    =
    \frac{1}{15}
    \sum_{k=1}^{15}
    \mathbb{I}(L_k=L_0),
\end{equation}
where $\mathbb{I}$ is the indicator function and $L_k$ is the label obtained from threshold combination $k$. A material retained its baseline assignment only when $R\geq0.80$, corresponding to agreement in at least 12 of the 15 threshold combinations. Configurations below this robustness criterion were labeled as ambiguous, and were excluded from the final binary FM/AFM classification problem.

Final task-level site moments were available for 118,510 of the 271,325 retrieved source calculations. Applying the above reconstruction and robustness procedure to this subset produced 92,933 NM, 17,189 FM, 2,792 FiM, 121 AFM, and 5,475 ambiguous configurations. Materials Project magnetic-order labels were not used directly as reconstructed target labels. They were retained for comparison and auditing and were also used to select AFM and FiM candidates for targeted recovery when final task-level moments were unavailable. This is important because a magnetic ordering associated with a single high-throughput calculation represents the spin configuration reached by that calculation and does not necessarily establish the global magnetic ground state unless competing magnetic configurations have been systematically explored \cite{horton2019high}. In particular, AFM order can be more difficult to represent in high-throughput calculations because it requires at least two oppositely polarized magnetic sublattices, and some ordering patterns require symmetry reduction or a magnetic supercell larger than the primitive crystallographic cell \cite{horton2019high}.

The remaining 152,815 source tasks did not contain a directly accessible final \texttt{magmom} array, although their initial moments and Materials Project processed magnetic information remained available. Because this missing-moment population contained 3,248 Materials Project AFM candidates and 10,318 FiM candidates, a targeted recovery step was performed to avoid disproportionately removing materials likely to contain antiparallel magnetic configurations. The recovery queue was therefore restricted to these AFM and FiM candidates, giving 13,566 materials in total. Their material records were queried through the Materials Project \texttt{materials/core} endpoint, and the site-resolved \texttt{magmom} values stored in the corresponding material structures were extracted. These recovered moments were processed using exactly the same active-site, alignment, sublattice-balance, and 15-combination robustness criteria described above.

Site-resolved moments were successfully recovered for all 13,566 materials in the targeted recovery set. Of these, 11,424 satisfied the $R\geq0.80$ robustness requirement, including 3,364 AFM, 6,838 FiM, 1,115 FM, and 107 NM assignments, while 2,142 configurations were classified as ambiguous. The recovery therefore increased the number of robust AFM configurations from only 121 in the directly available task-level dataset to a substantially larger population suitable for machine-learning analysis.

The two sources of magnetic evidence were subsequently combined using an explicit evidence hierarchy. Labels reconstructed directly from the final task-level magnetic moments were assigned evidence tier A, whereas labels recovered from material-structure magnetic moments were assigned evidence tier B. When duplicate material identifiers were encountered, entries were ordered first by evidence tier and then by label robustness, ensuring that a direct final-task result took precedence over a recovered structure-based result. After this consolidation, the curated dataset contained 132,076 unique materials, comprising 93,040 NM, 18,304 FM, 9,630 FiM, 3,485 AFM, and 7,617 ambiguous materials. Restricting the dataset to the two magnetic states considered in the present classification problem yielded 21,789 materials, consisting of 18,304 FM and 3,485 AFM compounds.

Following magnetic-state curation, an independent structure-only representation was constructed for machine learning, as summarized in Fig.~\ref{fig:1}b. This stage was implemented using \texttt{pymatgen} \cite{ong2013python}. To prevent magnetic information used for target construction from leaking into the input descriptors, each task structure was reconstructed after explicitly removing oxidation-state decorations and \emph{all} site properties, including the \texttt{magmom} values used during curation. The cleaned structure therefore retained only the lattice, atomic species, and fractional atomic coordinates. Magnetic moments, Materials Project magnetic-order labels, robustness values, evidence tiers, and all other magnetic quantities were explicitly prohibited from entering the model feature matrix.

The cleaned structures were converted into fixed-length descriptors describing composition, elemental properties, global crystal geometry, and local atomic environments. For each element $e$ in a material, the atomic fraction was defined as

\begin{equation}
    x_e = \frac{N_e}{N_{\mathrm{tot}}},
\end{equation}
where $N_e$ is the number of atoms of element $e$ and $N_{\mathrm{tot}}$ is the total number of atoms. In addition to the individual elemental fractions, the compositional diversity was quantified using the composition entropy

\begin{equation}
    S_{\mathrm{comp}}
    =
    -\sum_e x_e\ln x_e .
\end{equation}

For each elemental property $p_e$, composition-weighted statistics were calculated according to

\begin{equation}
    \bar{p}
    =
    \sum_e x_e p_e,
\end{equation}
and

\begin{equation}
    \sigma_p
    =
    \sqrt{
    \sum_e x_e
    \left(p_e-\bar{p}\right)^2
    }.
\end{equation}

The minimum and maximum values among the constituent elements were also retained. These statistics were calculated for atomic number, atomic mass, atomic radius, electronegativity, periodic-table row and group, and Mendeleev number. In this manner, the feature representation contains information about both the average chemical character of the material and the chemical contrast between its constituent species.

Global structural descriptors included the number of sites, number of elemental species, unit-cell volume, volume per atom, density, lattice constants $(a,b,c)$, lattice angles $(\alpha,\beta,\gamma)$, and the lattice ratios $a/b$, $b/c$, and $a/c$. Local structural information was obtained using periodic-neighbor searches implemented in \texttt{pymatgen}. For each atomic site $i$, a coordination count within a fixed $5~\text{\AA}$ radius was defined as

\begin{equation}
    N_i^{(5\text{\AA})}
    =
    \sum_{j\neq i}
    \Theta
    \left(
    5~\text{\AA}-r_{ij}
    \right),
\end{equation}
where $r_{ij}$ is the interatomic separation and $\Theta$ is the Heaviside step function. The corresponding nearest-neighbor distance was

\begin{equation}
    d_i^{\mathrm{NN}}
    =
    \min_{j\neq i} r_{ij}.
\end{equation}

For both $N_i^{(5\text{\AA})}$ and $d_i^{\mathrm{NN}}$, the mean, standard deviation, minimum, and maximum over all atomic sites were included as descriptors. The initial representation therefore contained 169 features. Thirty-three elemental-fraction features were constant across the development population and were removed prior to model fitting, leaving 136 active descriptors: 85 elemental-fraction features, 28 elemental-property statistics, 15 global structural descriptors, and 8 local-geometry descriptors.

The distribution of the resulting FM/AFM feature space is visualized using principal-component analysis in Fig.~\ref{fig:1}c. The first and second principal components account for 12.8\% and 5.3\% of the total variance, respectively. FM and AFM compounds occupy substantially overlapping regions of this low-dimensional representation, rather than forming two trivially separable clusters. This behavior is physically reasonable because the preference for parallel or antiparallel magnetic alignment is not expected to be controlled by a single structural parameter. Instead, the magnetic state reflects the combined influence of elemental electronic configuration, magnetic-ion separation, coordination environment, bonding geometry, and the exchange pathways permitted by the crystal structure. The strong overlap in the first two principal components therefore motivates the use of a nonlinear classifier capable of identifying multivariable relationships in the full higher-dimensional descriptor space. For comparison, Fig.~\ref{fig:1}d shows the same type of low-dimensional representation for the dataset used in the magnetization-prediction task, with each material colored by its magnetization per formula unit. Unlike the binary FM/AFM labels in Fig.~\ref{fig:1}c, the magnetization target varies continuously across the projected feature space. No single region of the first two principal components corresponds exclusively to either low or high magnetization; instead, materials with different magnetization values remain broadly distributed and partially intermixed. This indicates that the magnitude of the magnetization, similar to the magnetic-order classification problem, is not governed by a single dominant structural or compositional coordinate. Rather, it emerges from the combined influence of multiple descriptors associated with elemental composition, local coordination, crystal geometry, and the distribution of magnetic species within the structure. The broad continuous variation observed in Fig.~\ref{fig:1}d therefore further supports the use of nonlinear machine-learning models for quantitative magnetic-property prediction.

\begin{figure}[!ht]
    \centering
    \includegraphics[width=\linewidth]{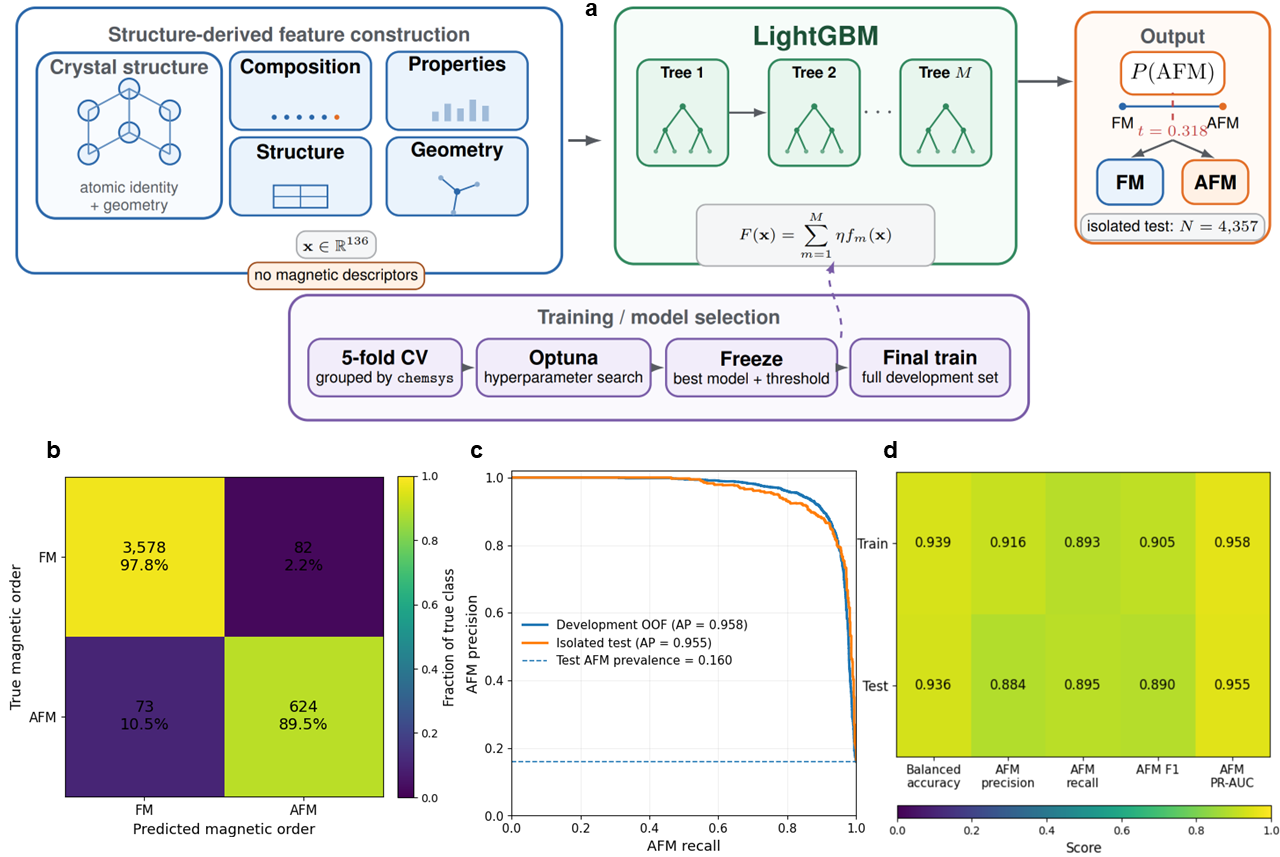}
    \caption{LightGBM framework and performance for FM/AFM magnetic-order classification.
    \textbf{(a)} Schematic of the magnetic-order classification pipeline. Structure-derived descriptors are constructed from crystal structure, composition, elemental properties, and geometric information, yielding a 136-dimensional feature vector without explicitly including magnetic descriptors. A LightGBM ensemble is optimized using five-fold cross-validation grouped by chemical system and Optuna hyperparameter search. The selected model and decision threshold are then frozen, retrained on the full development set, and evaluated on an isolated test set of 4,357 materials. The model outputs the probability of antiferromagnetic ordering, $P(\mathrm{AFM})$, which is converted to the final FM/AFM prediction using the fixed threshold $t=0.318$.
    \textbf{(b)} Confusion matrix for the isolated test set. The model correctly classifies 3,578 FM materials (97.8\%) and 624 AFM materials (89.5\%), while 82 FM and 73 AFM materials are misclassified.
    \textbf{(c)} Precision--recall curves for AFM classification obtained from out-of-fold predictions on the development set and predictions on the isolated test set. The corresponding average precision values are 0.958 and 0.955, respectively; the dashed line indicates the AFM prevalence of 0.160 in the test set.
    \textbf{(d)} Summary of classification performance on the training/development dataset and isolated test set, including balanced accuracy, AFM precision, AFM recall, AFM F1 score, and AFM precision--recall area under the curve (PR-AUC). The comparable development and test performance indicates that the optimized classifier generalizes consistently to previously unseen materials.}
    \label{fig:2}
\end{figure}

The final magnetic-order classification workflow is summarized in Fig.~\ref{fig:2}a. The 21,789-material FM/AFM dataset was separated into development and isolated test partitions according to chemical system. The development set contained 17,432 materials, consisting of 14,644 FM and 2,788 AFM compounds, while the locked test set contained 4,357 materials, including 3,660 FM and 697 AFM compounds. Materials belonging to the same chemical system were constrained to remain within the same partition. This strategy reduces the possibility that closely related chemical compositions appear in both the development and test sets and therefore provides a more stringent assessment of model transfer to previously unseen chemical systems.

The initial feature construction produced structural, compositional, elemental-property, and geometric descriptors. After descriptors with no variation across the development dataset were removed, 136 features were retained for model training. A LightGBM gradient-boosted decision-tree classifier \cite{ke2017lightgbm} was selected for the final classification task. During model development, five-fold cross-validation grouped by chemical system was used, and model hyperparameters were optimized using Optuna \cite{akiba2019optuna}. The probability threshold used to distinguish AFM from FM was determined using the development-set out-of-fold predictions and was subsequently frozen together with the model parameters before evaluation on the isolated test set. The final decision threshold was set to $P(\mathrm{AFM}) = 0.318$.

The confusion matrix for the isolated test set is shown in Fig.~\ref{fig:2}b. Of the 3,660 FM materials, 3,578 were correctly classified as FM and 82 were classified as AFM, corresponding to an FM recall of 97.8\%. For the AFM class, 624 of the 697 materials were correctly identified, while 73 were predicted as FM, yielding an AFM recall of 89.5\%. The ability to retain high AFM recall is particularly important given that AFM materials constitute only approximately 16\% of the dataset.

The precision--recall curves in Fig.~\ref{fig:2}c provide a complementary evaluation that is particularly informative for strongly imbalanced binary-classification problems \cite{saito2015precision}. The grouped out-of-fold development predictions achieved an AFM average precision of 0.958, while the isolated test set yielded an average precision of 0.955. For comparison, the AFM prevalence in the test set is approximately 0.160. The large separation between the model precision--recall curve and the prevalence baseline indicates substantial discriminative capability for the minority AFM class. In addition, the close agreement between the development and isolated-test curves indicates that the classification behavior is preserved when the model is applied to chemical systems excluded from model development.

The comparison of the principal classification metrics is shown in Fig.~\ref{fig:2}d. The development out-of-fold predictions achieved a balanced accuracy of 0.939, AFM precision of 0.916, AFM recall of 0.893, AFM $F_1$ score of 0.905, and AFM PR-AUC of 0.958. On the isolated test set, the corresponding values were 0.936, 0.884, 0.895, 0.890, and 0.955, respectively. The nearly unchanged AFM recall and PR-AUC between the development and isolated test sets indicate that the model retains its ability to distinguish AFM from FM configurations when evaluated on chemical systems not included during training.

These results demonstrate that FM and AFM configurations contain distinguishable statistical signatures in composition and crystal structure even when explicitly magnetic descriptors are excluded from the input representation. This result should not be interpreted as implying that magnetic order is determined by geometry independently of the electronic structure. Rather, the structural and compositional descriptors encode information correlated with the microscopic factors governing magnetic exchange, including elemental electronic configuration, magnetic-ion separation, coordination environment, bonding geometry, and possible exchange pathways. The nonlinear LightGBM model can combine these correlated descriptors to identify relationships between the structural environment and the resulting magnetic configuration, enabling magnetic-order classification without directly supplying magnetic quantities as model inputs.

\subsection*{Magnetization Value Prediction}

The regression task was constructed using the same 21,789-material FM/AFM population and the same frozen development/test partition used for magnetic-order classification, comprising 17,432 development materials and 4,357 chemically isolated test materials. The regression target was the Materials Project quantity \texttt{total\_magnetization\_normalized\_formula\_units}, corresponding to the total magnetic moment normalized by the number of formula units in the cell and reported in units of Bohr magnetons per formula unit, $\mu_{\mathrm{B}}$/f.u. All 21,789 materials contained a valid target value. The target distribution was strongly right-skewed, with a median magnetization of approximately $2.00~\mu_{\mathrm{B}}$/f.u. and values extending to approximately $98.1~\mu_{\mathrm{B}}$/f.u., motivating a regression strategy that explicitly accounts for the broad dynamic range of the target.

\begin{figure}
    \centering
    \includegraphics[width=\linewidth]{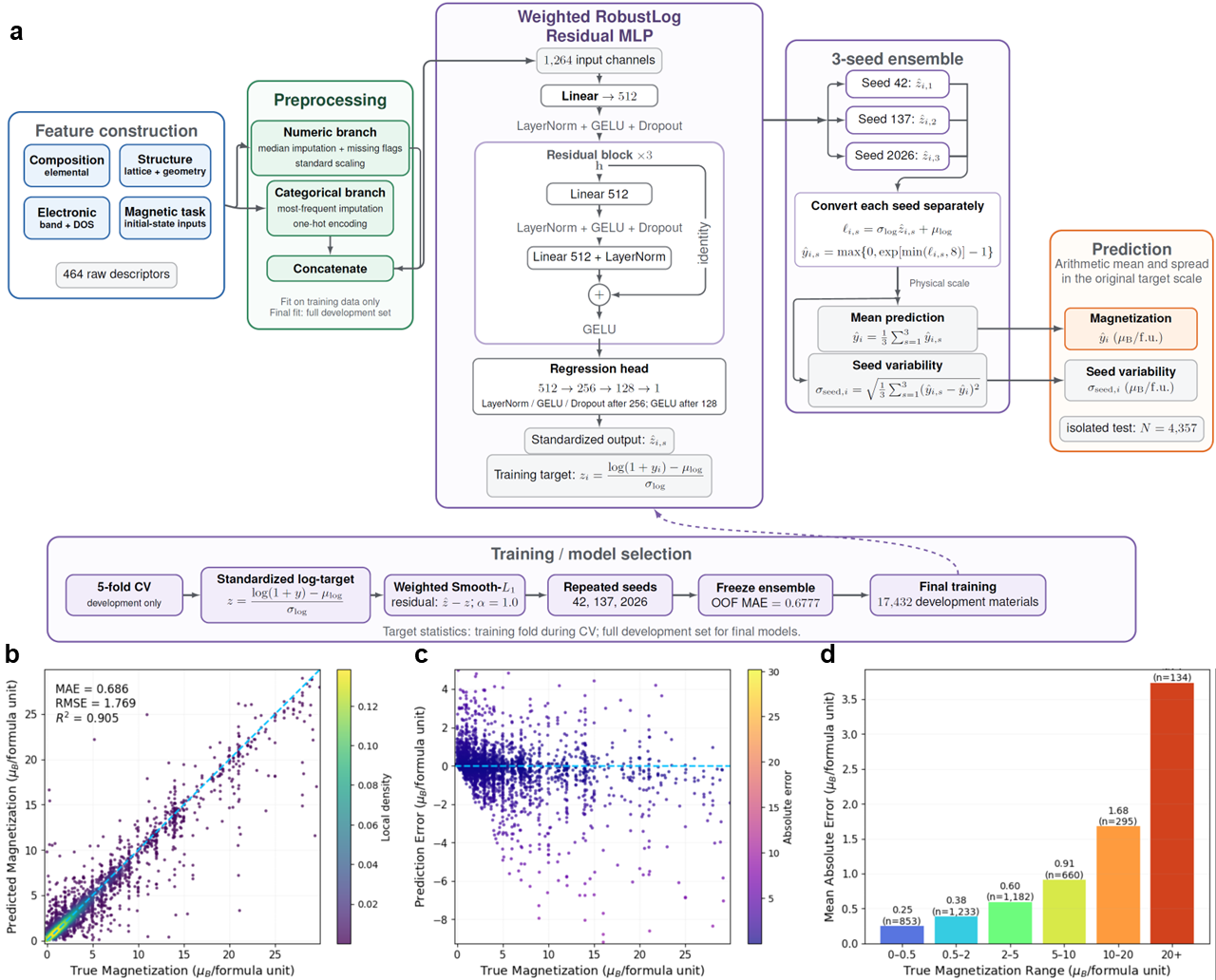}
    \caption{Residual multilayer perceptron framework and performance for magnetization prediction.
    \textbf{(a)} Schematic of the magnetization-regression pipeline. A 464-dimensional feature vector is constructed from compositional, structural, electronic, and magnetic-task initial-state information. Numerical features are median-imputed with missing-value indicators and standardized, while categorical features are imputed and one-hot encoded. The processed descriptors are passed to a weighted RobustLog residual multilayer perceptron (MLP) consisting of an initial 512-unit linear layer, three residual blocks, and a regression head with dimensions $512\rightarrow256\rightarrow128\rightarrow1$. The model is trained to predict the standardized log-target $z=[\log(1+y)-\mu_{\log}]/\sigma_{\log}$ using a weighted Smooth-$L_1$ loss with weighting strength $\alpha=1.0$. Model selection is performed using five-fold cross-validation on the development set and repeated training with three random seeds (42, 137, and 2026). The frozen three-seed ensemble is subsequently trained on the full development set of 17,432 materials. Final predictions are obtained by averaging the three model outputs and applying the inverse transformation, while the standard deviation across ensemble members, $\sigma_{\mathrm{seed}}$, provides an estimate of prediction variability. Performance is evaluated on an isolated test set of 4,357 materials.
    \textbf{(b)} Predicted versus true magnetization for the isolated test set, colored according to local point density. The dashed diagonal line indicates ideal agreement between prediction and ground truth. The model achieves a mean absolute error (MAE) of $0.686~\mu_{\mathrm{B}}$/formula unit, a root-mean-square error (RMSE) of $1.769~\mu_{\mathrm{B}}$/formula unit, and $R^{2}=0.905$.
    \textbf{(c)} Prediction residuals as a function of the true magnetization, where the residual is defined as the predicted value minus the true value and the color scale denotes the corresponding absolute error. The horizontal dashed line marks zero prediction error.
    \textbf{(d)} MAE evaluated over different ranges of true magnetization. The error increases from $0.25~\mu_{\mathrm{B}}$/formula unit for materials below $0.5~\mu_{\mathrm{B}}$/formula unit to approximately $3.7~\mu_{\mathrm{B}}$/formula unit for materials above $20~\mu_{\mathrm{B}}$/formula unit, illustrating the increasing prediction difficulty at large magnetization values.}
    \label{fig:3}
\end{figure}

In contrast to the magnetic-order classifier, which was intentionally restricted to composition and crystal-structure information, the magnetization-regression model used a broader physical representation, as illustrated in Fig.~\ref{fig:3}a. A total of 464 raw descriptors were retained, consisting of 457 numerical and 7 categorical features. Composition descriptors were generated using \texttt{matminer}, including Magpie elemental-property statistics, stoichiometric descriptors, average valence-orbital information, and elemental fractions \cite{ward2018matminer}. Structural descriptors were constructed from \texttt{pymatgen} structures and included lattice parameters, lattice angles, density, volume, volume per atom, number of sites, crystallographic symmetry, and lattice-shape ratios \cite{ong2013python}. Because the objective of this model is to reproduce the continuous magnetic response of an existing first-principles calculation rather than to infer magnetic order from structure alone, additional electronic and calculation-level information was also included.

The electronic representation contained band-gap and band-edge information together with descriptors extracted directly from the available band structures and densities of states. For a band energy $\varepsilon_{n\mathbf{k}}$ and Fermi energy $E_{\mathrm{F}}$, the energies were referenced to the Fermi level according to

\begin{equation}
    \Delta \varepsilon_{n\mathbf{k}}
    =
    \varepsilon_{n\mathbf{k}}-E_{\mathrm{F}},
\end{equation}
from which the energy range, number of bands crossing $E_{\mathrm{F}}$, and fractions of sampled band states lying within $\pm0.25$, $\pm0.50$, and $\pm1.00$~eV of the Fermi level were obtained. The density of states was similarly characterized by the total DOS at $E_{\mathrm{F}}$, integrated DOS within selected energy windows, and the position and magnitude of the dominant DOS peak within $\pm2$~eV. When spin-resolved DOS information was available, the Fermi-level spin polarization was calculated as

\begin{equation}
    P_{\mathrm{DOS}}(E_{\mathrm{F}})
    =
    \frac{
    \left|
    D_{\uparrow}(E_{\mathrm{F}})
    -
    D_{\downarrow}(E_{\mathrm{F}})
    \right|
    }
    {
    D_{\uparrow}(E_{\mathrm{F}})
    +
    D_{\downarrow}(E_{\mathrm{F}})
    +\epsilon
    },
\end{equation}
where $\epsilon=10^{-12}$ was included numerically to avoid division by zero. These descriptors provide information about the electronic states surrounding the Fermi level, where changes in occupation and spin polarization are closely connected to the resulting magnetic response.

Selected magnetic and calculation metadata were also included in the regression representation. These features included the magnetic ordering metadata, the numbers and identities of magnetic species, the fractions of magnetic and symmetry-unique magnetic sites, Hubbard-$U$ information, calculation type, run type, and the \emph{initial} magnetic moments supplied to the selected DFT task. For the initial moments $\{m_i^{(0)}\}$, quantities such as the signed, absolute, maximum, mean, and root-mean-square moments were retained, together with the initial alignment and sublattice-balance descriptors introduced during magnetic-state curation. Importantly, the final site-resolved magnetic moments and all quantities directly derived from the final magnetization target were explicitly excluded from the model inputs. Specifically, \texttt{magmoms}, \texttt{total\_magnetization}, \texttt{total\_magnetization\_normalized\_vol}, and the regression target itself were prohibited from entering the feature matrix. The regression therefore uses information available from composition, structure, electronic structure, materials-property metadata, and the initialization of the magnetic calculation, without directly supplying the final magnetization that the network is trained to predict.

The 464 raw descriptors exhibited substantially different numerical scales and levels of data coverage. Of the retained features, 359 were available for at least 95\% of the development materials, whereas 62 had intermediate coverage and 43 were sparsely available. The preprocessing procedure in Fig.~\ref{fig:3}a was therefore fitted separately within the development data. Missing numerical values were replaced by the median of the corresponding training feature, while an additional binary missingness indicator was introduced to preserve information about whether a value was originally available. Numerical variables were subsequently standardized as

\begin{equation}
    \tilde{x}_{ij}
    =
    \frac{x_{ij}-\mu_j}{\sigma_j},
\end{equation}
where $\mu_j$ and $\sigma_j$ were determined exclusively from the corresponding training data. Missing categorical values were imputed using the most frequent training category and were subsequently one-hot encoded. After numerical missingness indicators and categorical one-hot encoding were applied, the 464 raw descriptors expanded to 1,264 model-input channels for the final development set.

The wide range of magnetization values required additional treatment of the regression target. Direct optimization in the original target space places substantially greater numerical emphasis on the relatively small number of high-magnetization materials. The target was therefore transformed using a RobustLog representation,

\begin{equation}
    y_i^{\prime}
    =
    \log(1+y_i),
\end{equation}
where $y_i$ is the magnetization in $\mu_{\mathrm{B}}$/f.u. Because all targets were nonnegative, this transformation was well defined for the complete dataset. The transformed target was subsequently standardized,

\begin{equation}
    z_i
    =
    \frac{
    y_i^{\prime}-\mu_{\log}
    }
    {\sigma_{\log}},
\end{equation}
using the mean and standard deviation of $\log(1+y)$ calculated from the training population. The logarithmic transformation compresses the long high-magnetization tail while preserving the ordering of the target values, allowing the network to learn both low- and high-magnetization regimes on a more uniform numerical scale.

Although the logarithmic target improved the overall regression performance, it also reduces the relative contribution of the high-magnetization tail to the training loss. A target-magnitude-dependent weighting scheme was therefore introduced. For material $i$, an unnormalized sample weight was defined as

\begin{equation}
    w_i^{*}
    =
    1
    +
    \alpha
    \frac{\log(1+y_i)}
    {
    \left\langle
    \log(1+y)
    \right\rangle
    },
\end{equation}
followed by normalization to unit mean,

\begin{equation}
    w_i
    =
    \frac{w_i^{*}}
    {\left\langle w^{*}\right\rangle}.
\end{equation}

The weighting strength $\alpha$ was optimized using development-set cross-validation over
$\alpha=\{0,0.25,0.50,1.00\}$, with $\alpha=1.00$ selected for the final model. The resulting final-development weights ranged from approximately 0.50 to 2.43 while maintaining $\langle w\rangle=1$. This weighting increases the contribution of high-magnetization materials without discarding or undersampling the more numerous low- and intermediate-magnetization compounds.

The final model was a residual multilayer perceptron implemented in PyTorch \cite{paszke2019pytorch}. The network first projected the preprocessed input into a 512-dimensional latent representation, followed by three residual blocks based on identity skip connections \cite{he2016deep}. Within each residual block, the latent representation $\mathbf{h}_k$ was updated according to

\begin{equation}
    \mathbf{h}_{k+1}
    =
    \mathrm{GELU}
    \left[
    \mathbf{h}_{k}
    +
    \mathcal{F}_{k}
    \left(
    \mathbf{h}_{k}
    \right)
    \right],
\end{equation}
where

\begin{equation}
    \mathcal{F}_{k}
    =
    \mathrm{LN}
    \circ
    \mathrm{Linear}_{512}
    \circ
    \mathrm{Dropout}
    \circ
    \mathrm{GELU}
    \circ
    \mathrm{LN}
    \circ
    \mathrm{Linear}_{512}.
\end{equation}

Layer normalization was used to stabilize the hidden representations \cite{ba2016layer}, and Gaussian error linear unit activations were used throughout the nonlinear portions of the network \cite{hendrycks2016gaussian}. The residual representation was passed to a regression head with dimensions
$512\rightarrow256\rightarrow128\rightarrow1$, as shown in Fig.~\ref{fig:3}a.

Training used a weighted Smooth-$L_1$ objective, closely related to the Huber robust loss \cite{huber1992robust}. For a standardized-target residual
$r_i=\hat{z}_i-z_i$, the loss with $\beta=0.25$ was

\begin{equation}
    \ell_{\beta}(r_i)
    =
    \begin{cases}
    \dfrac{r_i^2}{2\beta},
        & |r_i|<\beta, \\[6pt]
    |r_i|-\dfrac{\beta}{2},
        & |r_i|\geq\beta,
    \end{cases}
\end{equation}
and the optimized batch objective was

\begin{equation}
    \mathcal{L}
    =
    \frac{1}{N}
    \sum_{i=1}^{N}
    w_i\,
    \ell_{\beta}
    \left(
    \hat{z}_i-z_i
    \right).
\end{equation}

This objective retains a quadratic response to small errors while reducing the influence of very large residuals compared with a purely squared-error loss. Optimization was performed using AdamW \cite{loshchilov2017decoupled}, with an initial learning rate of $8\times10^{-4}$, weight decay of $3\times10^{-4}$, a batch size of 256, dropout of 0.15, and gradient clipping at a norm of 5.

Model development was performed entirely on the 17,432-material development set. Five-fold cross-validation was constructed by jointly stratifying the data by the curated FM/AFM label and by five within-class magnetization-rank intervals. This procedure maintained both the magnetic-order composition and the target distribution across the validation folds. Initial tabular regression models, including LightGBM, XGBoost, CatBoost, and Extra Trees, were evaluated under the same development framework as discussed in Supplementary Information \ref{sec:Mag_Regress_Compar}. The best of all models trained, Extra Trees, achieved an out-of-fold MAE of $0.837~\mu_{\mathrm{B}}$/f.u. A residual MLP substantially reduced this error, and incorporation of the RobustLog target and target-magnitude weighting further improved the development predictions.

Because neural-network optimization can vary with random initialization, the final configuration was repeated using seeds 42, 137, and 2026. For $\alpha=1.00$, the three individual repeated-seed runs produced a mean out-of-fold MAE of approximately $0.694~\mu_{\mathrm{B}}$/f.u. Averaging their predictions further reduced the development out-of-fold MAE to $0.6777~\mu_{\mathrm{B}}$/f.u., and the three-seed ensemble was therefore frozen before the isolated test set was evaluated. The final training epoch counts for seeds 42, 137, and 2026 were 82, 99, and 111, respectively, obtained from the median best epoch across the corresponding five development folds. Each network was subsequently retrained using all 17,432 development materials.

For each material, the three independently trained models produced predictions $\hat{z}_{i,s}$, with $s$ denoting the random seed. Each prediction was returned to the physical target space according to

\begin{equation}
    \hat{y}_{i,s}=\max\left\{0,\exp\!\left[\min\!\left(\sigma_{\log}\hat{z}_{i,s}+\mu_{\log},8\right)\right]-1\right\}.
\end{equation}

with predicted log-value was capped at 8 before exponentiation, and negative magnetization predictions clipped to zero. The final ensemble prediction was the arithmetic mean,

\begin{equation}
    \hat{y}_{i}
    =
    \frac{1}{3}
    \sum_{s=1}^{3}
    \hat{y}_{i,s},
\end{equation}
and the seed-to-seed disagreement was retained as

\begin{equation}
    \sigma_{\mathrm{seed},i}
    =
    \sqrt{
    \frac{1}{3}
    \sum_{s=1}^{3}
    \left(
    \hat{y}_{i,s}
    -
    \hat{y}_{i}
    \right)^2
    }.
\end{equation}

Here, $\sigma_{\mathrm{seed}}$ should be interpreted as an empirical measure of model disagreement under repeated initialization rather than as a formally calibrated predictive uncertainty.

The final frozen ensemble was evaluated once on the 4,357-material isolated test set. As shown in Fig.~\ref{fig:3}b, the predicted magnetizations follow the ideal $\hat{y}=y$ relationship over a broad range of target values. The model achieved a mean absolute error of
$0.686~\mu_{\mathrm{B}}$/f.u., a root-mean-square error of
$1.769~\mu_{\mathrm{B}}$/f.u., and a coefficient of determination of
$R^2=0.905$. The high-density region is concentrated close to the diagonal, indicating that the majority of low- and intermediate-magnetization materials are reproduced with comparatively small absolute errors. The larger deviations are concentrated primarily among the more sparsely represented high-magnetization compounds.

This behavior is more clearly visible in the residual distribution in Fig.~\ref{fig:3}c, where the prediction error is defined as

\begin{equation}
    \Delta M
    =
    \hat{M}-M.
\end{equation}

For the densely populated low-magnetization region, the residuals remain concentrated near zero. With increasing true magnetization, however, the residual distribution broadens and develops a tendency toward negative values, indicating increasing underprediction of the upper tail. This behavior is consistent with the strongly imbalanced continuous target distribution: although target weighting increases the contribution of large-magnetization materials during training, such materials remain substantially less numerous than compounds in the central portion of the target distribution.

The dependence of the absolute error on the true magnetization is quantified in Fig.~\ref{fig:3}d. For the 853 materials with magnetizations below $0.5~\mu_{\mathrm{B}}$/f.u., the MAE is approximately $0.25~\mu_{\mathrm{B}}$/f.u. The error increases to approximately $0.38$, $0.60$, and $0.91~\mu_{\mathrm{B}}$/f.u. for the $0.5$--$2$, $2$--$5$, and $5$--$10~\mu_{\mathrm{B}}$/f.u. intervals, containing 1,233, 1,182, and 660 materials, respectively. For the 295 materials between $10$ and $20~\mu_{\mathrm{B}}$/f.u., the MAE increases to approximately $1.68~\mu_{\mathrm{B}}$/f.u., while the 134 materials above $20~\mu_{\mathrm{B}}$/f.u. exhibit the largest errors, with an MAE of approximately $3.7~\mu_{\mathrm{B}}$/f.u. The monotonic increase in absolute error therefore primarily reflects the increasingly sparse and numerically broad high-magnetization regime rather than a uniform degradation across the dataset.

Thus the isolated-test results show that the regression framework captures the continuous magnetization response over a wide range of magnetic materials while retaining transfer to chemical systems excluded from model development. The remaining error is concentrated primarily in the sparsely sampled high-magnetization tail, suggesting that further improvement will depend particularly on increasing the representation of high-moment materials or introducing representations that more explicitly resolve the electronic mechanisms governing large spin polarization.

\subsection*{Magnetic Image Processing and Topological Order Identification}
We use the Single-Image Transport of Intensity Equation (SITIE) analysis to analyze the vector fields with phase retrieved from the simulated Lorentz transmission electron microscopy (LTEM) image using \textit{PyLorentz} python package. \cite{mccray2021understanding}
Under the paraxial approximation, the electron wavefunction can be expressed as \cite{mccray2021understanding}
\begin{equation}
\Psi(x,y,z) = A(x,y,z) \, e^{i \phi(x,y,z)},
\end{equation}
where $\phi_t$ is the phase shift of the electron wave, and 
$A(x,y,z)$ is the amplitude function by the shape properties. In the \textit{PyLorentz} package, the amplitude function is fined as:

\begin{equation}
A(x,y,z) = e^{-t/\xi_0},
\end{equation}

Therefore, the image intensity is given by $I(x,y,z) = |A(x,y,z)|^2$.  
The TIE establishes a direct relationship between the derivative of the image intensity along the optical axis $z$ and the divergence of the transverse phase gradient\cite{mccray2021understanding}:
\begin{equation}
\frac{\partial I(x,y,z)}{\partial z} 
= -\frac{\lambda}{2\pi} \nabla_{\perp} \cdot 
\left[ I(x,y,z) \, \nabla_{\perp} \phi_t(x,y,z) \right],
\label{eq:TIE}
\end{equation}
where $\lambda$ is the relativistic electron wavelength and $\nabla_{\perp}$ denotes the gradient in the image plane $(x,y)$.

\textit{PyLorentz} also implement the feature of Tikhonov regularization within the Fourier transform of the Laplacian operator $\frac{1}{q^2}$ as such:

\begin{equation}
\frac{q^2}{(q^2+q_c^2)^2}
\end{equation}
where $q_c$ is the Tikhonov frequency. The Tikhonov regulator can reduce the low frequency noise and the high frequency noise, especially when the image is spanned over a large field with a large amount of noise leading to a relatively low quality of the magnetic phase image. A band-pass filter can be effectively implemented with \textit{PyLorentz} to pre-process the image and achieve a better SITIE analysis results in better image quality. 

Additionally, to map the HSV color space into a 3D vector space, we are looking to analyze the image color using a lightweight convolution-neural-network called TinyCNN consisting of two convolution layers with rectified linear unit (ReLU) activation, an adaptive average pooling layer, and two fully connected layers projecting to $\mathbb{R}^3$.  
Given an input tile $\mathbf{C}_i \in [0,1]^{t \times t \times 3}$, the network outputs a vector $\hat{\mathbf{m}}_i \in \mathbb{R}^3$ which is normalized:
\begin{equation}
\hat{\mathbf{m}}_i = \frac{f_{\theta}(\mathbf{C}_i)}{\| f_{\theta}(\mathbf{C}_i) \|}.
\end{equation}

with mean squared vector error:
\begin{equation}
\mathcal{L}(\theta) = \frac{1}{N} \sum_{i=1}^N \| \hat{\mathbf{m}}_i - \mathbf{m}^{\mathrm{true}}_i \|_2^2.
\end{equation}

\begin{figure}[ht]
    \centering
    \includegraphics[width=0.9\textwidth]{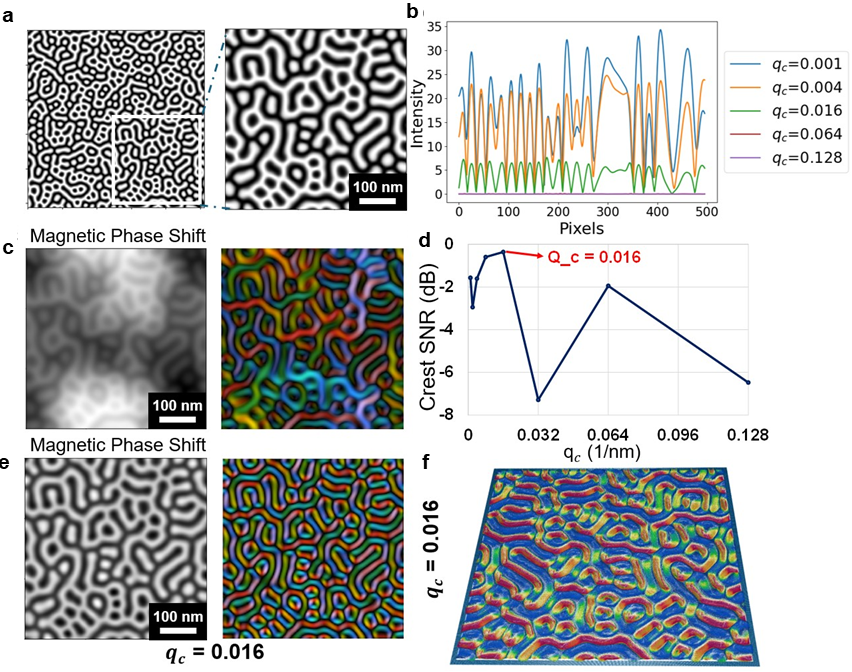}
    \caption{Magnetic image processing and phase-reconstruction results for skyrmion-like texture identification. 
    \textbf{(a)} Simulated magnetic image and enlarged region of interest showing labyrinth-like magnetic contrast. 
    \textbf{(b)} Intensity line profiles extracted under different cutoff wave-vector values $q_c$, showing how filtering affects image contrast and spatial oscillation. 
    \textbf{(c)} Reconstructed magnetic phase shift before cutoff optimization, shown as grayscale phase contrast and corresponding color phase map. 
    \textbf{(d)} Crest signal-to-noise ratio (SNR) as a function of cutoff wave vector $q_c$, where the optimal value is identified as $q_c = 0.016~\mathrm{nm}^{-1}$. 
    \textbf{(e)} Magnetic phase-shift reconstruction using the selected cutoff value, showing improved spatial contrast and clearer magnetic-domain features. 
    \textbf{(f)} Three-dimensional visualization of the reconstructed magnetic phase texture at $q_c = 0.016~\mathrm{nm}^{-1}$.}
    \label{fig:4}
\end{figure}

We took a simulated LTEM image of Fe$_3$GaTe$_2$, and performed the SITIE analysis on the image that includes the mixture of skyrmion domain and stripe domain as shown in (Fig. \ref{fig:4}a) From the image with the scale of 1024 $\times$ 1024, we took a 500 $\times$ 500 section from the lower right corner. As previously discussed, the Tikhonov filter would improve magnetic phase shift image quality. Therefore, we evaluate the performance of different Tikhonov frequency value $q_c$ as a high pass filter. We first take the simulation image without added noise as shown in Fig. \ref{fig:4}a. We perform SITIE analysis with $q_c$ ranging from $0.001$ to $0.128$ in geometric ascending order to find the optimum Tikhonov frequency value. The induction line profile of the magnetic phase texture intensity across the x-axis for each filtered image was shown in Figure \ref{fig:4}b. The expectation of the optimum Tikhonov frequency is to provide the best possible magnetic phase structure with very little noise. As shown in Figure \ref{fig:4}c, when no Tikhonov filter is applied, the analyzed magnetic phase image is unable to reproduce the true magnetic texture. From Figure \ref{fig:4}b, we noticed the average intensity of the image is reduced. Therefore, we calculate the crest signal-to-noise (SNR) for each profile line and found that $q_c = 0.016$ gives the best SNR as shown in Fig. \ref{fig:4}d. The Tikhonov filter played a significant role on improving the magnetic phase image quality. When applying the Tikhonov filter with $q_c = 0.016$ nm$^{-1}$, the magnetic phase image is improved. From the reconstructed magnetic phase image, we were able to reconstruct the 3D visualization of the magnetic as shown in Fig. \ref{fig:4}f.

\section*{Discussion}
Two primary prediction tasks were developed: magnetic-order classification and quantitative magnetization prediction. For magnetic-order classification, a LightGBM model was trained using 136 composition- and structure-derived descriptors while explicitly excluding magnetic quantities from the input representation. Using a chemically isolated locked test set of 4,357 materials, the classifier achieved an overall accuracy of approximately 0.964 and a balanced accuracy of 0.936. For the minority AFM class, the model achieved a precision of 0.884, recall of 0.895, $F_1$ score of 0.890, and PR-AUC of 0.955. The close agreement between the development out-of-fold and isolated-test performance indicates that the model retains substantial predictive capability when applied to chemical systems that were not included during model development.

The classification results suggest that FM and AFM configurations contain distinguishable statistical signatures in composition and crystal structure, even when descriptors that explicitly encode magnetic information are excluded. This should not be interpreted as implying that magnetic order is determined independently of the electronic structure. Instead, the structural and compositional descriptors serve as indirect representations of the microscopic factors governing magnetic exchange, including elemental electronic configuration, interatomic separation, local coordination, bonding geometry, and possible exchange pathways. The strong overlap between FM and AFM materials in the low-dimensional PCA representation further shows that the distinction cannot be attributed to a single dominant descriptor. The performance of the nonlinear LightGBM model therefore suggests that useful information about magnetic ordering is distributed across combinations of structural and chemical variables.

The magnetization-prediction task was formulated separately because predicting a continuous magnetic response requires a richer representation than binary magnetic-order classification. The regression model used composition, crystal-structure, electronic band-structure, density-of-states, magnetic-metadata, and magnetic-task initialization information, while explicitly excluding the final magnetic moments and quantities directly derived from the target magnetization. The final model consisted of a weighted RobustLog residual multilayer perceptron evaluated as a three-seed ensemble. On the same chemically isolated test set of 4,357 materials, the model achieved a mean absolute error of $0.686~\mu_{\mathrm{B}}/\mathrm{f.u.}$, a root-mean-square error of $1.769~\mu_{\mathrm{B}}/\mathrm{f.u.}$, and an $R^2$ value of 0.905. The close correspondence between predicted and reference magnetization over most of the target range indicates that the combined structural and electronic representation captures a substantial fraction of the variation in the calculated magnetic response.

The remaining regression error is not distributed uniformly across the magnetization range. Materials with magnetization below $0.5~\mu_{\mathrm{B}}/\mathrm{f.u.}$ exhibit an MAE of approximately $0.25~\mu_{\mathrm{B}}/\mathrm{f.u.}$, whereas the error progressively increases for larger target values, reaching approximately $3.7~\mu_{\mathrm{B}}/\mathrm{f.u.}$ for materials above $20~\mu_{\mathrm{B}}/\mathrm{f.u.}$. The residual distribution similarly broadens at high magnetization and shows a tendency toward underprediction. This behavior is consistent with the strongly imbalanced continuous target distribution, in which high-magnetization compounds constitute only a small fraction of the available data. The RobustLog transformation and target-dependent weighting reduce this imbalance during optimization but cannot fully compensate for the limited number and greater physical diversity of materials in the high-magnetization tail. Expanding this region of the training dataset is therefore likely to be particularly important for further improving the quantitative prediction of large magnetic moments.

An important aspect of the present work is the physics-informed reconstruction and curation of the magnetic targets. Rather than directly adopting the categorical magnetic-order labels provided by the Materials Project, the magnetic states were reconstructed from site-resolved local moments using the degree of magnetic compensation and the balance between oppositely oriented magnetic sublattices. A robustness analysis across multiple active-moment and alignment thresholds was additionally used to remove assignments that depended sensitively on a particular numerical cutoff. This procedure substantially increased the number of usable AFM configurations through the recovery of magnetic information from materials for which final task-level moments were unavailable. Nevertheless, the resulting dataset remains strongly imbalanced toward FM materials. This imbalance reflects both the available database population and the greater difficulty of systematically representing competing AFM configurations in high-throughput calculations.

Several limitations should therefore be considered when interpreting the present results. First, the magnetic-order targets represent magnetic configurations reconstructed from available DFT calculations and should not automatically be interpreted as experimentally confirmed magnetic ground states. Establishing a true magnetic ground state generally requires comparison between multiple competing magnetic configurations, and some AFM states may require symmetry lowering or magnetic supercells that are not represented in the available calculation. Second, the regression target is a calculated magnetization at the DFT level rather than a temperature-dependent experimental magnetization. The present model consequently learns relationships associated with the available zero-temperature or ground-state-like first-principles data and does not describe the thermal evolution of the magnetic state.

The combination of magnetic-order classification and magnetization prediction nevertheless provides a useful first-stage screening framework. The classification model can identify materials whose structural and compositional characteristics are associated with FM or AFM configurations, while the regression model provides an estimate of the magnitude of the corresponding magnetic moment in units of $\mu_{\mathrm{B}}/\mathrm{f.u.}$. Used together, the two tasks can act as an effective screening process narrow down to reduced number of material candidates. In particular, the strong isolated-test performance suggests that the framework can be used to prioritize chemically distinct candidates rather than only interpolating among closely related compounds already present in the training data.

Future development should incorporate finite-temperature magnetic behavior into this screening framework. A particularly important target is the prediction of magnetic transition temperatures, including the Curie temperature, $T_{\mathrm{C}}$, for ferromagnetic materials and the N\'eel temperature, $T_{\mathrm{N}}$, for antiferromagnetic materials. Incorporating these quantities would allow the workflow to move beyond identifying magnetic ground-state configurations and magnetization values toward evaluating whether a predicted magnetic phase can remain stable at technologically relevant temperatures. Such an extension would be especially valuable for the search for robust room-temperature van der Waals antiferromagnets, for which both the magnetic ordering and the corresponding transition temperature are essential screening criteria.

The broader framework can also be extended from material-level magnetic-property prediction to the identification and reconstruction of spatially varying magnetic textures. Machine-learning-assisted magnetic-image analysis provides a possible route for detecting textures such as magnetic skyrmions after candidate materials have been identified through the preceding screening stages. However, an image-based reconstruction is fundamentally constrained by the information accessible from the observed surface or projection. Future work should therefore incorporate physics-informed models capable of constraining the magnetic configuration throughout the material thickness. Reconstruction of the full three-dimensional spin texture would make it possible to investigate whether surface-observed skyrmions extend through the sample as continuous skyrmion tubes or exhibit depth-dependent deformation, termination, or discontinuities. Such information would provide a more complete description of the three-dimensional topology of these magnetic structures and could ultimately help connect materials screening, magnetic-texture detection, and the assessment of topological stability within a unified machine-learning-assisted workflow.

\section*{Methods}
\subsection*{Model Selection and Training Criteria}
Different training and model-selection strategies were adopted for the magnetic-order classification and magnetization-regression tasks because the two problems have different statistical objectives, target distributions, and feature representations. For the FM/AFM classification task, the primary challenge is the strong class imbalance between ferromagnetic and antiferromagnetic materials. Consequently, overall classification balanced accuracy alone was not considered an appropriate model-selection criterion, since a model biased toward the majority FM class could achieve a high apparent accuracy while performing poorly on AFM materials. Instead, the classifier models were developed and trained using five-fold cross-validation grouped by chemical system, such that chemically related materials were kept within the same fold to reduce information leakage between structurally or compositionally similar compounds and to provide a more realistic estimate of generalization to previously unseen chemical systems. Hyperparameters were optimized on the development data, and the probability threshold used to convert the predicted AFM probability, $P(\mathrm{AFM})$, into the final FM/AFM label was optimized independently for each model.

Model selection for the classification task was therefore based on a combination of balanced accuracy, AFM precision, AFM recall, AFM F1 score, macro F1 score, AFM precision--recall area under the curve (PR-AUC), FM recall, and ROC-AUC rather than on a single metric. Particular emphasis was placed on AFM-related metrics because AFM materials is the minority class and are consequently more difficult to achieve a relatively high accuracy without overfitting the classifier to all AFM prediction. Among all tested algorithms, as shown in Table \ref{tab:1}, LightGBM provided one of the strongest overall performance profiles. At an optimized threshold of $0.318$, LightGBM achieved a balanced accuracy of $0.9390$, AFM precision of $0.9161$, AFM recall of $0.8935$, AFM F1 score of $0.9047$, macro F1 score of $0.9434$, AFM PR-AUC of $0.9581$, and ROC-AUC of $0.9833$. Although XGBoost produced a slightly higher balanced accuracy ($0.9416$) and AFM recall ($0.9014$), LightGBM achieved higher in other metrics measuring the generalization ability and was able to provided a more balanced overall trade-off between minority-class identification and global classification performance.

The magnetization-prediction problem required a different training criterion because it is a continuous regression task rather than a discrete classification problem. In addition, the magnetization values span a broad numerical range and the distribution contains a relatively small number of materials with very large magnetization. Direct optimization in the original target space would therefore cause large-magnetization samples and large residuals to contribute disproportionately to the loss. To reduce this effect, the target was transformed using
\begin{equation}
    y^{'}=\log(1+y),    
\end{equation}
which compresses the dynamic range while preserving the ordering of the target values. A weighted robust-loss formulation was then used to further reduce the sensitivity of the model to unusually large residuals. In contrast to the classification task, where tree-based ensemble methods are well suited to the lower-dimensional tabular descriptor space, the regression model uses a substantially larger $464$-dimensional representation containing compositional, structural, electronic, and magnetic initial-state information. A residual multilayer perceptron was therefore employed to learn nonlinear interactions across this higher-dimensional feature space.

For the regression task, model selection was based primarily on out-of-fold regression error rather than classification-oriented metrics. Five-fold cross-validation was performed using only the development data, and the model configuration was evaluated using the mean absolute error (MAE) of the out-of-fold predictions. Because neural-network training can vary with random initialization and stochastic optimization, the final model was trained repeatedly using three fixed random seeds ($42$, $137$, and $2026$). Each ensemble member was subsequently trained on the full development set of $17{,}432$ materials. For the isolated test set, the final prediction was obtained from the mean of the three independently trained models,
\begin{equation}
    \hat{y}_{i,s}=\max\left\{0,\exp\!\left[\min\!\left(\sigma_{\log}\hat{z}_{i,s}+\mu_{\log},8\right)\right]-1\right\},
\end{equation}
where $\mu_{\log}$ and $\sigma_{\log}$ were calculated from the log-transformed targets of the full development set. The predicted log-value was capped at 8 before exponentiation, and negative predictions were clipped to zero. The final magnetization prediction was obtained by averaging the three converted predictions:
\begin{equation}
    \hat{M}_i\equiv\hat{y}_i=\frac{1}{3}\sum_{s=1}^{3}\hat{y}_{i,s}.
\end{equation}
The standard deviation among the three seed predictions, $\sigma_{\mathrm{seed}}$, was additionally retained as a measure of prediction variability. Thus, the classification pipeline was designed primarily around class imbalance, minority-class sensitivity, and threshold optimization, whereas the regression pipeline was designed around continuous-error minimization, robustness to the broad magnetization distribution, and stability with respect to neural-network initialization. In both tasks, the isolated test set was kept completely separate from model selection and was used only after the final model configuration had been frozen.

\begin{table*}[!ht]
\centering
\caption{\textbf{Comparison of tuned machine-learning models for FM/AFM magnetic-order classification.}}
\resizebox{\textwidth}{!}{%
\begin{tabular}{lccccccccc}
\toprule
\textbf{Model} &
\textbf{Threshold} &
\textbf{Balanced Accuracy} &
\textbf{AFM Precision} &
\textbf{AFM Recall} &
\textbf{AFM F1} &
\textbf{FM Recall} &
\textbf{Macro F1} &
\textbf{AFM PR-AUC} &
\textbf{ROC-AUC} \\
\midrule

LightGBM
& 0.318 & 0.9390 & 0.9161 & 0.8935 & 0.9047 & 0.9844 & 0.9434 & 0.9581 & 0.9833 \\

XGBoost
& 0.420 & 0.9416 & 0.9046 & 0.9014 & 0.9030 & 0.9819 & 0.9423 & 0.9547 & 0.9826 \\

CatBoost
& 0.443 & 0.9366 & 0.9033 & 0.8913 & 0.8973 & 0.9818 & 0.9389 & 0.9523 & 0.9828 \\

HistGradientBoosting
& 0.520 & 0.9317 & 0.9020 & 0.8816 & 0.8917 & 0.9818 & 0.9357 & 0.9510 & 0.9820 \\

ExtraTrees
& 0.440 & 0.9293 & 0.8859 & 0.8802 & 0.8831 & 0.9784 & 0.9304 & 0.9400 & 0.9799 \\

RandomForest
& 0.483 & 0.9314 & 0.8606 & 0.8902 & 0.8752 & 0.9725 & 0.9255 & 0.9357 & 0.9782 \\

SVM-RBF
& 0.284 & 0.9165 & 0.8646 & 0.8587 & 0.8616 & 0.9744 & 0.9177 & 0.9220 & 0.9687 \\

Logistic Regression
& 0.592 & 0.9018 & 0.8441 & 0.8329 & 0.8384 & 0.9707 & 0.9039 & 0.8774 & 0.9616 \\

\bottomrule
\label{tab:1}
\end{tabular}%
}
\end{table*}

\subsection*{Micromagnetic Simulation}
MuMax$^3$ \cite{vansteenkiste2014design} with Dzyaloshinskii–Moriya interaction (DMI) module was used to perform micromagnetic simulations. The governing equation of the simulation is based on the Landau-Lifshitz-Gilbert (LLG) equation \cite{gilbert2004phenomenological}: 
\begin{equation}
\
\frac{\partial\bm{M}}{\partial t}
  = \gamma^{*}\,\bm{M}\!\times\!\bm{H}
  - \frac{\alpha}{M_s}\,\bm{M}\!\times\!\frac{\partial\bm{M}}{\partial t},
\
\label{eq:LLG}
\end{equation}
where,
\begin{equation}
\gamma^{*} = \gamma(1+\alpha^2).
\end{equation}

The term $\gamma$ is the gyromagnetic ratio, closely linked to $\alpha = 0.3$, the Gilbert damping factor.  The term \(M_{\rm S}\) is the saturation magnetization. In this simulation, we take the following paramters for Fe$_3$GaTe$_2$ \cite{lv2024distinct}: \(M_{\rm S} =\) $2.4\times 10^4$ A/m, $A_{\rm ex} = 5\times10^{-12}$ J/m, DMI $=0.8\times 10^{-3}$ J/m$^2$, Temp $=313$ K, $K_{\rm anis} = 8\times 10^4$ J/m$^3$. We simulated the material at the size of 1024 $\times$ 1024 $\times$ 10 with the cell size of 2 nm $\times$2 nm $\times$2 nm, and only the magnetization vector respective to z-axis are kept to make a grazy scale image to simulate a magnetic single image. Our workflow took a 500 $\times$ 500 section from the lower right corner. In practice, the model can be adjust to process magnetic texture image under any size.

\printbibliography

\section*{Data Availability}

The code are available in UF QESI webpage. 

\section*{Acknowledgment}

We thank Dr. Chunjing Jia for the helpful discussion. The work of Yingying Wu was supported in part by National Science Foundation DMR grant No. 2501208. 

\section*{Competing Interests}
Author Y. Wu is Editorial Board Member of \textit{npj Spintronics}. Y. Wu was not involved in the journal’s review of, or decisions related to, this manuscript. The remaining authors declare no competing financial or non-financial interests.

\section*{Author Contributions}
Y. Wu conceived the ideas and supervised the project. H. Yang conducted model training and development, with S. Huerta. All authors contributed to writing and revising the manuscript.

\end{document}